\documentclass[useAMS,usenatbib,usegraphicx]{mn2e}
\usepackage{multirow}
\usepackage{url}
\usepackage{rotating}
\title[Counterparts for \emph{Fermi}/LAT 1$^{st}$ Catalogue Objects] 
{Using the \emph{Rosat} Catalogue to find Counterparts for Unidentified Objects in the 1$^{st}$ \emph{Fermi}/LAT Catalogue}

\author[J. B. Stephen et al.]
{J.~B.~Stephen,$^1$\thanks{E-mail: stephen@iasfbo.inaf.it}  L.~Bassani,$^1$ R.~Landi,$^1$ A.~Malizia,$^1$
V.~Sguera,$^1$ A.~Bazzano,$^2$ \newauthor and N.~Masetti$^1$\\
$^1$INAF/IASF-Bologna, Via P. Gobetti 101, I-40129 Bologna, Italy \\
$^2$INAF/IASF-Roma, Via Fosso del Cavaliere 100, I-00133, Roma, Italy}
\begin{document}
         
\date{Accepted 2010 June 01. Received 2010 June 01; in original form 2010 April 28}

\pagerange{\pageref{firstpage}--\pageref{lastpage}} \pubyear{2010}

\maketitle 

\label{firstpage}

\begin{abstract}
There are a total of 1451 gamma-ray emitting objects in the \emph{Fermi} Large Area Telescope First Source Catalogue. 
The point source location accuracy of typically a few arcminutes has allowed the counterparts for many of these sources
to be found at other wavelengths, but even so there are 630 which are described as having no plausible counterpart at $80\%$ confidence.
In order to help identify the unknown objects, we have cross-correlated the positions of these sources with the \emph{Rosat} All Sky Survey Bright Source Catalogue. In this way, for \emph{Fermi} sources which have a possible counterpart in soft X-rays, we can use the, much smaller, \emph{Rosat} error box to search for identifications. We find a strong correlation between the two samples and calculate that there are about 60 sources with a \emph{Rosat} counterpart. Using the \emph{Rosat} error boxes we provide tentative associations for half of them, demonstrate that the majority of these are either blazars or blazar candidates and give evidence that most belong to the BL Lac class. Given that they are X-ray selected and most are high synchrotron peaked objects, which indicates the presence of high energy electrons, these sources are also good candidates for TeV emission, and therefore good probes of the extragalactic background light.
\end{abstract}


\section{Introduction} 

A key strategic objective of the \emph{Fermi} mission is a survey of the sky at gamma-ray energies, making use of the large area and field of view of the LAT instrument \citep{atwood}. The telescope allows the detection of sources with an angular resolution of about 0.6 degrees (68\% at 1 GeV) and a point source location accuracy (PSLA) varying from around 1 to 6 arcminutess, depending on the detection significance. In the first \emph{Fermi} catalogue (\citealt{abdo1}), comprising data from the initial 11 months of the science phase of the mission, there are 1451 objects listed, of which 821 have been associated with known sources at other wavelengths. These identified sources comprise both extragalactic and galactic objects with the former including blazars (flat spectrum radio QSOs (FSRQ) or BL Lacs), a few radio galaxies and 4 normal galaxies, while the latter is made up of pulsars, pulsar wind nebulae, supernova remnants, globular clusters and a few binaries. Some peculiar objects are also found but their associations are less secure.  No plausible counterparts have been found for the remaining 630 objects and therefore these cannot yet be associated with any known class of gamma-ray emitting objects. The \emph{Fermi}/LAT AGN catalogue (\citealt{abdo2}, hereafter A1) lowers the confidence limit to 50\% for high latitude ($|b| > 10^{\circ}$) sources  and thereby provides possible counterparts for another 26 of these sources, and for 104 more there are potential associations ('affiliations') but with unquantified confidence.

Searching for counterparts of these new high energy sources is a primary objective of the survey work but it is made 
very difficult by the large, with respect to other wavelengths, \emph{Fermi} error boxes. This uncertainty in their locations  means that a positional correlation with a known object is usually not enough to identify a \emph{Fermi} source and instead, a multiwavelength approach, using X-ray, optical and radio data of likely counterparts must be used in order to understand their nature and to evaluate the likelihood of their association with the \emph{Fermi} detections. Searches for X-ray counterparts are particularly useful in finding a positionally-correlated, highly-unusual object with the special parameters that might be expected to produce gamma rays. X-ray surveys are well suited for this type of search because they offer 3 great advantages: a) they allow a full coverage of the \emph{Fermi} error box, b) they provide arcsecond location accuracy and c) they give information in an energy band quite close to that in which \emph{Fermi} operates. Cross correlation analysis using X-ray catalogues can therefore be a useful tool with which to restrict the positional uncertainty of the objects detected by \emph{Fermi} and so to facilitate the identification process.

Herein, we report on the strong level of positional correlation between the unassociated \emph{Fermi} sources and the \emph{Rosat} All Sky Survey Bright Source Catalogue  leading to evidence for the association of a number of GeV sources with a soft X-ray counterpart, better positions for all correlated objects and hence the possibility of optical follow-up work. We find that most of these associations are with BL Lac or BL Lac candidates. 

\section{The cross-correlation}

The \emph{Rosat} all-sky survey was  performed in the period July 1990 to February 1991  with the X-ray telescope (XRT) and the Position Sensitive Proportional Counter \citep{pepper}. The survey mapped 145,060 sources in the soft X-ray band (0.1-2.4 keV) from which the Bright Source Catalogue (RASSBSC-1.4.2RXS), containing 18806 RASS sources having a PSPC count rate larger than 0.05 cts/s and at least 15 source counts, was extracted  \citep{voges}. Many of these sources have been identified with objects expected to emit even at high energies, i.e in the \emph{Fermi}/LAT regime. 

To perform the correlation, instead of using the entire first year \emph{Fermi} catalogue, we chose to use only those 630 which were listed as unassociated in order to keep the possibility of chance positional coincidences to a minimum. We used the standard statistical technique which has been employed very succesfully to help identify sources found in the various
\emph{INTEGRAL}/IBIS surveys (\citealt{stephen1}; \citealt{stephen2}; \citealt{stephen3}). This consists of simply calculating the number of \emph{Fermi} sources for which at least one
\emph{Rosat} counterpart was within a specified distance, out to a distance where all \emph{Fermi} sources had at least one \emph{Rosat} counterpart.
To have a control group we create a list of fake 'anti-\emph{Fermi}' sources.  For every object in the \emph{Fermi} list we make a corresponding source in the fake list with coordinates mirrored in Galactic longitude and latitude (this mirroring was chosen due to the strong galactic component evident in the LAT distribution), and the same correlation algorithm was then applied betweeen this list and the \emph{Rosat} catalogue. Figure 1a shows the results of this process.
The lower solid curve is the 'anti-source' correlation, while the dashed line is that expected from chance correlations given the number of \emph{Fermi} objects and the number of \emph{Rosat} sources. It is clear that, in this case, the number of correlations can be 
completely explained by chance. The upper solid curve, however, shows the number of associations for the \emph{Fermi} unidentified 
sources, and demonstrates that a strong correlation exists. 

\begin{figure}
\includegraphics[width=8.4cm]{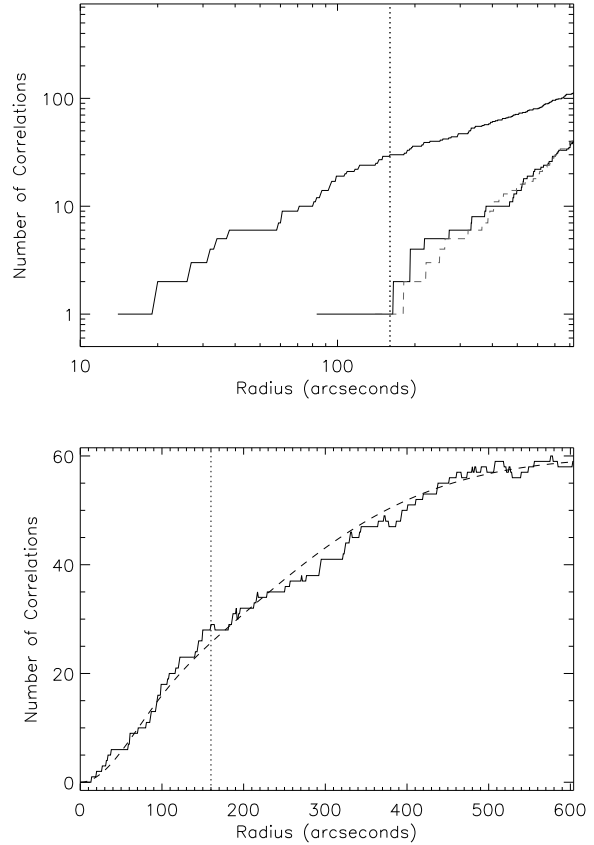}
\caption{a \emph{top}: The number of  LAT/\emph{Rosat} (upper solid) and 'Anti-LAT'/\emph{Rosat} (lower solid) associations as a function of distance. The dashed line shows the number of correlations expected by chance.
b: The difference between the number of true and false associations. The dashed line is the sum of several inverse Gaussian functions having widths between 1 and 4 arcminutes. In both figures the maximum correlation distance considered is indicated by the vertical dotted line}
\end{figure}

Figure 1b shows the difference in the number of associations between the true and false data sets. The point at which the curve flattens off gives the total number of associations present in the data set, which is around 60. In this analysis we have not used the positional errors of either catalogue, but the shape of the curve should be compatible with the PSLA of \emph{Fermi} as the uncertainy in the \emph{Rosat} position is negligible in comparison. By fitting a series of inverse gaussians to the data we find the curve to be consistent with a combination of location errors varying between around 1 and 4 arcminutes (dashed line in Figure 1b), consistent with the range in errors quoted for the \emph{Fermi} sources.

Clearly, even though there are around 60 true associations in the dataset, we cannot search for counterparts for them all as we do not know exactly which sources are correctly identified as being correlated. At around 500 arcseconds, all we know is that around 60 of the 77 correlations found will be correct. For this reason we limit the correlation distance to 160 arcseconds, where less than two false associations are expected, from a total of 30 sources with possible RASSBSC counterparts. In order to provide a rough estimate of the probability of association we note that the number of false sources is expected to be 1$\pm$1. This indicates that, given no other information, the likelihood of any one source being associated correctly is around 74\%. If we can assume that the extra information on X-ray source type allows us to assert that we know the identity of one chance alignment (1FGL J0841.4-3558, see below) then this probability will rise to over 90\% for any individual source in the list.

\section{Searching for counterparts of unidentified \emph{Fermi} sources} 

For the 30 \emph{Fermi}  sources which have  a possible counterpart within 160 arcseconds, Table 1
reports the \emph{Fermi} name, \emph{Rosat} coordinates, both instruments' error box radius as well as the
distance of the \emph{Rosat} position from that of the \emph{Fermi}/LAT location. The \emph{Rosat} uncertainty  provides a smaller 
error box than \emph{Fermi}  by about an order of magnitude, and so allows an easier  search for counterparts; furthermore in many cases XMMSLew \citep{saxton}, XMM serendipitous \citep{watson}
and/or\emph{Swift}/XRT measurements\footnote{The \emph{Swift}/XRT data analysis is performed using the standard procedure, see \citet{landi}
for details} provide an even smaller error box and therefore an unambigous source identification.  In the following we discuss
each individual \emph{Rosat} source in detail trying to provide  when possible an indication of its nature and class 
(see last two columns in table 1).  In particular, we use X-ray and radio data from the  literature and/or 
archives to assess the object type and hence its association with the \emph{Fermi} detection.

Historically AGN were discovered by radio observations. Radio emission is often a way to 
recognize active galaxies, except at lower luminosities where star-formation in galaxies can also stimulate radio 
production. Therefore, for bright objects, mere detection in radio provides support for the presence of an 
active galaxy, although contamination from Galactic sources may come from SNR, pulsars and microquasars.
In some cases, the radio spectrum, morphology and loudness can help in discriminating between the above 
possibilities since a compact  source with a flat spectrum which is radio loud is often indicative of a blazar type AGN,
i.e. those strongly correlated with emission  in the GeV domain. So, while mere radio detection does not imply identification with an 
AGN, the combination of X (and even more gamma-ray) emission 
plus association with a loud, compact and flat spectrum radio source provides strong support (augmented if it is located 
away from the galactic plane) for the extragalactic nature of an unclassified object  and further 
suggests a Blazar classification.

For this study, we inspect radio images  taken from the NVSS (NRAO VLA Sky Survey, \citealt{condon}) and the SUMSS surveys (Sydney University Molonglo Sky Survey, \citealt{mauch}). All sources which have a radio flux listed in Table 1, have a compact 
radio structure except for the case of the supernova remnant  G043.3-00.2. 
A flat radio spectrum is  often indicative of a Blazar type object: indeed in A1
the overall distribution is consistent with a flat spectral index
($\alpha_r= 0.08\pm0.32$). No difference is found between FSRQs and BL Lacs.
Information on the  radio spectrum of our sources (see Table 2), can be obtained from SpecFind \citep{vollmer} which is a tool used to cross-identify radio sources in various 
catalogues on the basis of self-consistent spectral index as well as position. This allows the combination of 
data at different frequencies and the estimation of the source radio spectrum as $Log(S(\nu)) = a \times Log(\nu) 
+ b$ where $S$ in expressed in Jy and $\nu$ in MHz. Flat spectrum sources are those with a$\ge$ --0.5
Only four (1FGL J0137.8+5814, 1FGL J2056.7+4938, 1FGL J2329.2+3755 and 
1FGL J1926.8+6153 ) of the 30 \emph{Rosat} sources had information about the spectral slope in the Specfind database and all were found to have a flat radio spectrum. 
Indication of a flat spectrum in other sources can be found in the literature (\citealt{reich}; \citealt{ribo}; \citealt{tsarevsky}; \citealt{jackson}; \citealt{mahony1}; \citealt{landi}) or 
from archival radio data 
(taken from HEASARC and/or NED); when available a radio loudness indication is also reported in the discussion 
of each individual source.
The \emph{Rosat} flux, calculated in the 0.1-2.4 keV band, has also been estimated and listed in table 2.
 
As a final point, we note that 14 objects listed in the tables appear  in A1 and 
all are listed  as affiliations. These are AGN or AGN candidates found inside the 95$\%$ \emph{Fermi} error ellipse that show hint of blazar
characteristics such as a radio and X-ray emission plus indication in the literature of variability polarization etc.
For 9  of these  objects a Spectral Energy Distribution (SED) class is also reported; this is based on a scheme discussed in A1. 
According to this scheme all 9  are  High Synchrotron Peaked (HSP) AGN, a   type of object
almost invariably associated to BL Lacs in A1.
Indeed in 3 cases (1FGL J1926.8+6153,  1FGL J2042.2+2427 and 1FGL J1553.5-3116) an optical classification as BL Lac is also provided by the \emph{Fermi} team.

\begin{table*}
\begin{center}
\caption{Unidentified \emph{Fermi} Sources with a possible RASSBSC counterpart: \emph{Rosat} position, identification and class}
\begin{tabular}{lcccccc}
\hline
\hline
\multicolumn{1}{|c|}{\emph{Fermi} Name}      & \emph{Rosat}           & \multicolumn{2}{|c|}{$^\star$Error($'$)} & Distance& ID (NED)       &class \\ 
&Coordinates&\emph{Rosat} & \emph{Fermi}& ($'$)&&\\
\hline
1FGL J1942.7$+$1033$^\dagger$     & 19 42 46.3 $+$10 33 39.0 & 0.23  &1.7 & 0.25  & 87GB 194024.3$+$102612     & BL Lac \\
1FGL J1307.6$-$4259                & 13 07 37.8 $-$42 59 40.5 & 0.17  & 2.0 & 0.30  & 1RXS J130737.8$-$425940    & BL Lac?      \\      
1FGL J0648.8$+$1516$^\dagger$     & 06 48 47.8 $+$15 16 26.0 &  0.12 &1.3 & 0.47  & 2MASX J06484763$+$1516248  & Blazar\\
1FGL J1353.6$-$6640$^\dagger$     & 13 53 41.1 $-$66 40 02.0 &  0.13 & 2.0  & 0.52  & VASC J1353$-$66            & BL Lac? \\
1FGL J0137.8$+$5814$^\dagger$     & 01 37 48.0 $+$58 14 22.5 & 0.15  & 4.0 & 0.58  & 87GB 013433.2$+$575900     & BL Lac   \\ 
1FGL J0604.2$-$4817                & 06 04 09.4 $-$48 17 26.5 & 0.17 & 2.3 & 0.67  & 1RXS J060409.4$-$481726    & BL Lac              \\
1FGL J0506.9$-$5435                & 05 06 56.8 $-$54 34 56.5 &  0.13 & 2.1  & 0.98  & RBS 621                  & BL Lac\\
1FGL J1304.3$-$4352                & 13 04 21.2 $-$43 53 08.0 & 0.15  & 1.6 & 1.02  & 1RXS J130421.2$-$435308    & AGN? \\
1FGL J1823.5$-$3454$^\dagger$     & 18 23 39.2 $-$34 54 12.5 & 0.12  &2.3& 1.02  & NVSS J182338$-$345412      & AGN?\\
1FGL J1227.9$-$4852                & 12 27 58.8 $-$48 53 43.5 &  0.13  &3.6&  1.22  & XSS J12270$-$4859          & Binary \\
1FGL J1643.5$-$0646                & 16 43 28.1 $-$06 46 27.0 & 0.28  &2.8& 1.33  & 2MASX J16432892$-$0646190  & BL Lac? \\
1FGL J0838.6$-$2828$^\dagger$      & 08 38 42.1 $-$28 27 23.0 & 0.25  &5.8& 1.42  & 1RXS J083842.1$-$282723    & ?            \\
1FGL J0051.4$-$6242$^\dagger$      & 00 51 17.7 $-$62 41 54.0 & 0.20  &2.5& 1.43  & RBS 119                  & BL Lac \\
1FGL J0131.2$+$6121$^\dagger$      & 01 31 06.4 $+$61 20 35.0 &  0.12  &1.3& 1.43  & 87GB 012752.4$+$610507     & BL Lac \\
1FGL J2056.7$+$4938$^\dagger$      & 20 56 44.3 $+$49 40 11.5 & 0.27  &1.6& 1.57  & MG4 J205647$+$4938         & Blazar/$\mu$QSO \\
1FGL J2146.6$-$1345                & 21 46 37.3 $-$13 43 55.5 & 0.17  &2.6&  1.60  & NVSS J214637$-$134359      & BL Lac?\\
1FGL J1544.5$-$1127                & 15 44 39.4 $-$11 28 20.5 &  0.13  &6.7& 1.60  & 1RXS J154439.4$-$112820    & ?  \\
1FGL J0848.6$+$0504                & 08 48 40.1 $+$05 06 30.5 & 0.23  &4.8& 1.68  & SDSS J084840.20$+$050611.9 & AGN       \\
1FGL J2329.2$+$3755                & 23 29 14.2 $+$37 54 15.0 & 0.15  &1.1& 1.68  & NVSS J232914$+$375414      & BL Lac?\\
1FGL J1926.8$+$6153                & 19 26 49.5 $+$61 54 45.0 &  0.12  &2.1& 1.80  & 87GB 192614.4$+$614823     & BL Lac  \\
1FGL J2042.2$+$2427                & 20 42 06.3 $+$24 26 55.5 &  0.12  &3.5&  1.85  & 2MASX J20420606$+$2426518  & BL Lac \\
1FGL J0841.4$-$3558$^\dagger$      & 08 41 21.4 $-$35 57 04.5 &  0.15  &3.2& 1.95  & HIP 42640                & Star \\
1FGL J1910.9$+$0906$^\dagger$     & 19 11 07.4 $+$09 06 23.5 & 0.37  &0.8 &  1.98  & SNR G043.3$-$00.2          & SNR \\
1FGL J0054.9$-$2455                & 00 54 47.2 $-$24 55 32.0 &  0.20 &2.4& 2.02  & NVSS J005446$-$245529      & BL Lac?     \\
1FGL J1933.3$+$0723$^\dagger$      & 19 33 20.3 $+$07 26 16.0 &  0.22 &2.9& 2.38  & 1RXS J193320.3$+$072616    & AGN?      \\
1FGL J1553.5$-$3116                & 15 53 33.4 $-$31 18 41.0 &  0.15  &2.4& 2.38  & 1RXS J155333.4$-$311841    & BL Lac   \\     
1FGL J1841.9$+$3220                & 18 41 47.0 $+$32 18 38.5 &  0.13  &2.7& 2.40  & RGB J1841$+$323            & BL Lac?\\
1FGL J1419.7$+$7731                & 14 19 01.8 $+$77 32 29.0 & 0.15  &2.6& 2.48  & 1RXS J141901.8$+$773229    & AGN?   \\
1FGL J2323.0$-$4919                & 23 22 56.7 $-$49 16 58.0 & 0.27  &3.5& 2.50  & $-$$-$$-$$-$                & AGN?    \\
1FGL J0223.0$-$1118                & 02 23 14.6 $-$11 17 41.0 & 0.20  &3.6&  2.67  & NVSS J022314$-$111737      & AGN       \\             

\hline 
\hline
\end{tabular}
\end{center}
$^\star$The \emph{Fermi} error is the average of the semi-major and -minor axes at 68\%while the \emph{Rosat} error is the 1-$\sigma$ radius\\
$\dagger$ object at low galactic latitude, i.e. within $\pm$10 degrees of the galactic plane\\
\end{table*}

\begin{table*}
\begin{center}
\caption{The gamma-ray flux and spectral slope, X-ray and radio fluxes, curvature and variability indices for the correlated sources } 
\begin{tabular}{lccccccc}
\hline
\hline
\multicolumn{1}{|c|}{\emph{Fermi} Name}&1-100 GeV Flux& \emph{Fermi} Photon & 0.1-2.4 keV Flux$^\dagger$ & 20/36 cm Flux$^\ddagger$& $\alpha_R$ & CI     & VI \\ 
           &10$^{-9}$ ph cm$^{-2}$ s$^{-1}$&  index   & 10$^{-12}$ erg cm$^{-2}$ s$^{-1}$ &  (mJy) &                       &        &\\
\hline

1FGL J1942.7$+$1033                & 3.40$\pm$0.47   & 1.77$\pm$0.10        &   1.38$\pm$0.26  &  99$^a$    & $+$0.10   &  11.99 & 8.3     \\
1FGL J1307.6$-$4259$^\star$    & 1.74$\pm$0.36  & 1.89$\pm$0.14        &   1.84$\pm$0.26  &  36$^b$    & $-$$-$$-$     &  0.78  & 11.8    \\     
1FGL J0648.8$+$1516                 & 1.76$\pm$0.33  & 1.81$\pm$0.11        &   7.37$\pm$0.43  &  64$^a$    & $+$0.01   &  1.09  & 7.1\\
1FGL J1353.6$-$6640                  & 1.51$\pm$0.41  & 2.34$\pm$0.18        &   1.74$\pm$0.20  &  $-$$-$$-$        & $-$0.20   &  2.66  & 4.2      \\
1FGL J0137.8$+$5814                 & 1.42$\pm$0.40  & 2.39$\pm$0.17        &   2.92$\pm$0.42  &  170$^a$   & $-$0.30   &  4.08  & 5.2 \\              
1FGL J0604.2$-$4817$^\star$    & 1.12$\pm$0.31  & 2.11$\pm$0.16        &   2.58$\pm$0.50  &  32$^b$    &  $-$$-$$-$    &  2.81  & 6.7 \\  
1FGL J0506.9$-$5435$^\star$    & \hspace{-15bp}0.98$^+$          & 1.42$\pm$0.31        &   5.44$\pm$0.58  &  18$^b$    &  $-$$-$$-$    &  0.35  &  4.0  \\
1FGL J1304.3$-$4352$^\star$    & 3.69$\pm$0.48  & 2.05$\pm$0.08        &   1.72$\pm$0.29  &  44$^a$    &  $-$$-$$-$    &  1.18  &  8.6  \\
1FGL J1823.5$-$3454                  & 1.78$\pm$0.39  & 1.70$\pm$0.12        & \hspace{-7bp} 17.35$\pm$1.18  &  132$^a$   & $-$0.24   &  3.00  &  9.3  \\
1FGL J1227.9$-$4852                  & 3.41$\pm$0.49  & 2.45$\pm$0.07        &   3.39$\pm$0.32  &  $-$$-$$-$        &  $-$$-$$-$    &  8.41  &  3.7  \\
1FGL J1643.5$-$0646$^\star$    & 2.34$\pm$0.44 & 2.21$\pm$0.10        &   0.67$\pm$0.13  &  28$^a$    &  $-$$-$$-$    &  0.89  & 12.2  \\
1FGL J0838.6$-$2828                  & 1.21$\pm$0.32  & 2.12$\pm$0.14        &   0.76$\pm$0.17  &  $-$$-$$-$        &  $-$$-$$-$    &  8.05  & 12.0  \\
1FGL J0051.4$-$6242$^\star$    & 1.84$\pm$0.32  & 1.68$\pm$0.12        &   2.89$\pm$0.43  &  43$^b$    &  $-$$-$$-$    &  4.00  &  9.3   \\
1FGL J0131.2$+$6121                 & 3.58$\pm$0.52  & 2.27$\pm$0.08        &   2.50$\pm$0.24  &  19$^a$    & $-$0.27   &  4.00  &  7.2   \\
1FGL J2056.7$+$4938                 & 1.67$\pm$0.62  & 1.85$\pm$0.19        &   0.82$\pm$0.15  &  167$^a$   & $+$0.40   &  1.68  &  4.3  \\
1FGL J2146.6$-$1345$^\star$    & 1.48$\pm$0.30  & 1.82$\pm$0.16        &   0.85$\pm$0.18  &  22$^a$    &  $-$$-$$-$    &  0.64  & 10.4    \\
1FGL J1544.5$-$1127                  & 1.35$\pm$0.35  & 2.45$\pm$0.14        &   1.04$\pm$0.20  &  $-$$-$$-$        &  $-$$-$$-$    &  5.79  &  9.8   \\
1FGL J0848.6$+$0504                 &\hspace{-15bp}0.76$^+$          & 1.24$\pm$0.35        &   2.03$\pm$0.30  &  2$^a$     &  $-$$-$$-$    &  0.52  & 1.9       \\
1FGL J2329.2$+$3755$^\star$   & 1.28$\pm$0.29  & 1.61$\pm$0.17        &   0.97$\pm$0.19  &  20$^a$    & $+$0.36   &  2.82  &  5.7   \\
1FGL J1926.8$+$6153$^\star$   & 1.94$\pm$0.32  & 2.02$\pm$0.10        &   2.03$\pm$0.30   &  22$^a$    & $-$0.26   &  5.13  &  9.0     \\
1FGL J2042.2$+$2427$^\star$   & 0.99$\pm$0.29  & 1.92$\pm$0.18        &   5.38$\pm$0.32  &  70$^a$    & $-$0.24   &  1.17  &  6.5     \\
1FGL J0841.4$-$3558                  & 1.31$\pm$0.35  & 1.76$\pm$0.20        &   1.46$\pm$0.24  &  $-$$-$$-$        &  $-$$-$$-$    &  1.20  &  10.5  \\
1FGL J1910.9$+$0906                 & \hspace{-4bp}24.13$\pm$1.45 & 2.23$\pm$0.03        &   1.51$\pm$0.21  &  8146$^a$  &  $-$$-$$-$    &  8.59  &  12.1 \\
1FGL J0054.9$-$2455$^\star$    & 0.73$\pm$0.22  & 1.95$\pm$0.22        &   1.83$\pm$0.26  & 24.1$^a$     &         & 1.58   & 9.87 \\
1FGL J1933.3$+$0723                 & 0.99$\pm$0.40  & 2.32$\pm$0.17        &   0.95$\pm$0.19  & 94.3$^a$     &  $-$0.08  & 3.15   & 6.46 \\
1FGL J1553.5$-$3116$^\star$    & 1.12$\pm$0.35  & 1.67$\pm$0.15        &   1.03$\pm$0.18  & 155.6$^a$    &  $+$0.19  & 1.85   & 5.52  \\
1FGL J1841.9$+$3220$^\star$   & 1.40$\pm$0.35  & 2.14$\pm$0.14        &   2.06$\pm$0.18  &  20.4$^a$    &  $+$0.30  &  0.43  & 12.6     \\
1FGL J1419.7$+$7731                 & 0.50$\pm$0.23  & 1.88$\pm$0.29        &   1.09$\pm$0.19  &  8.1$^a$  &  $-$$-$$-$    &  0.66  & 6.1  \\
1FGL J2323.0$-$4919$^\star$    &  \hspace{-15bp}0.83$^+$          & 1.62$\pm$0.28        &   1.96$\pm$0.38  &  23.8$^b$ &  $-$$-$$-$    &  1.85  & 2.0 \\
1FGL J0223.0$-$1118                  &\hspace{-15bp}0.86$^+$          & 1.50$\pm$0.29        &   0.51$\pm$0.13  &  14$^a$    &  $-$$-$$-$    &  0.37  & 5.1   \\
\hline
\hline
\end{tabular}
$^\star$Affiliated AGN in the First catalogue of AGN detected by the \emph{Fermi} Large Area telescope \citep{abdo2}\\
$^+$2$\sigma$ upper limit\\
$\dagger$ assuming a flux conversion factor of 1PSPC count = 10$^{-11}$ erg cm$^{-2}$ s$^{-1}$  \\
$\ddagger$ a: 20cm  flux from  NVSS \citep{condon}; b 36 cm flux from SUMSS \citep{mauch} \\
\end{center}
\end{table*}

\section{The Individual Sources}

In the following, we provide detailed information on each individual source as available in the literature and in various archives, in the same order as in the tables (by correlation distance).

{\bf 1FGL J1942.7+1033}. This \emph{Rosat} source is identified with a radio object having very similar flux at 20 and 6 cm (around 100 mJy) and so is likely 
to be a flat spectrum source \citep{tsarevsky}. The optical spectrum is featureless which, when combined with the detection at radio and  X-ray frequencies, suggests that  it is probably  a new BL Lac type object located behind the plane of the Galaxy  and it is classified as such in NED.

{\bf 1FGL J1307.6-4259}. This X-ray emitter is still unidentified. The only secure radio detection is at 36 cm with a  flux of 
36 mJy; however it is possible that PMN J1307-4259  at a distance of 0\arcmin.34 is also  
a radio counterpart of  1RXS J130737.8-425940 in which case its radio flux at around 6 cm is 53 mJy \citep{griffith1}, again indicative of a flat spectrum source. The source location above the galactic plane, its radio and X-ray emission all suggest
that this is most likely an AGN; indeed it is listed as an affiliated source in the \emph{Fermi} first AGN catalogue
with a HSP SED class and hence could be a BL Lac candidate.

{\bf 1FGL J0648.8+1516}. This \emph{Rosat} detection  is also reported in the  first XMM-Newton slew survey catalogue as 
XMMSL1 J064847.6+151626; the smaller XMM Slew error box allows the unambigous identification of this 
(and the \emph{Rosat}) source with the galaxy 2MASX J06484763+1516248 which is still unclassified in NED.
This galaxy is radio detected at 6 and 20 cm with a flux of 67 and 64 mJy (see NED photometric database) and has a flat radio spectrum \citep{mahony1}; the source was also reported as a radio loud  AGN  first  by \citet{brinkman} and later by \citet{laurent}.
The 0.2-12 keV flux  is 6.5$\times$10$^{-12}$ erg cm$^{-2}$ s$^{-1}$. Recently the source has been observed with VERITAS and found to be 
a source of very high energy photons \citep{ong}. There are 4 \emph{Swift}/XRT observations of this source: the spectrum from each each can be fitted with an absorbed power-law with purely galactic N$_{H}$ giving a photon index which ranges from 2.15 to 2.78 (typical error $\pm$0.1). The 0.2-12 keV flux also varies within a wide range (2.16 - 0.76$\times$10$^{-11}$ erg cm$^{-2}$ s$^{-1}$. All these properties suggest that the X-ray source is 
the  likely extragalactic counterpart of the \emph{Fermi} object, probably a blazar at low galactic latitudes.

{\bf 1FGL J1353.6-6640}. This source also has  a counterpart in the XMMSlew catalogue (XMMSL1 J135340.5-663958) which provides a restricted error
 box, a secure identification with VASC J1353-66  and a 0.2-12 keV flux of 3.9$\times$10$^{-12}$ erg cm$^{-2}$ s$^{-1}$. \citet{tsarevsky} report that this source has a featureless optical spectrum and a flat radio spectrum which together with the detected X-ray and radio emission leads these authors to suggest that it is probably 
a new BL Lac type object behind our galaxy despite the  marginal detection of a small proper motion.
 
{\bf 1FGL J0137.8+5814}. The \emph{Rosat} position is compatible with an XMM source (2XMM J013750.3+581410) serendipitously detected in the field of PSR B0136+57
which is aound 11 arcminutes away. It has a 0.2-12 keV flux of 1.8$\times$10$^{-12}$ erg cm$^{-2}$ s$^{-1}$. This source is also 
associated with an \emph{INTEGRAL} object first reported by \citet{krivonos} in their all-sky hard 
X-ray survey and also listed in the fourth IBIS catalogue \citep{bird}. It is relatively bright in radio, 
it has a flat spectrum between 6 and 82 cm \citep{vollmer} and is radio loud (\citealt{brinkman}; \citealt{laurent}). 
The optical spectrum is featureless indicating that this is another BL Lac object \citep{bikmaev}.

{\bf 1FGL J0604.2-4817}. This source has also a counterpart in the XMMSlew catalogue (XMMSl1 J060408.5-481712) which provides a much better position
and a 0.2-12 keV flux of 7.1$\times$10$^{-12}$ erg cm$^{-2}$ s$^{-1}$. In this case the source has also been observed by\emph{Swift}/XRT; the observed  spectrum
is a  simple power law absorbed by the galactic  column density (N$_H$=3.64$\times$10$^{20}$ cm$^{-2}$) and having 
a photon index $\Gamma$ = 2.3$\pm0.1$; the   0.2-12 keV flux of 5.6$\times$10$^{-12}$ erg cm$^{-2}$ s$^{-1}$
is  slightly lower than that measured by 
XMM during slews, suggesting  variable X-ray emission. The source is located at high galactic latitude and has a  counterpart 
in the SUMSS catalogue with a 36 cm flux of 32 mJy. 
All these properties suggest an  AGN nature. Indeed the source appears in the \emph{Fermi} AGN catalogue as an affiliated source with no further information, however recently it has been classified as a BL Lac on the basis of its optical spectrum \citep{mahony2}.

{\bf 1FGL J0506.9-5435}. This \emph{Rosat} source has also an  XMM Slew counterpart (XMMSL1 J050658.2-543503) which is associated 
to the source RBS 621,
classified in NED and Simbad as a BL Lac object; the source is also a \emph{Fermi} AGN affliation although with no optical nor SED class. 
RBS 621 is fairly bright in X-rays with a 0.1-12 keV flux of 
9.5$\times$10$^{-12}$ erg cm$^{-2}$ s$^{-1}$ while the only radio detection of the source is at 36 cm with a relatively low flux of 
18 mJy. 

{\bf 1FGL J1304.3-4352}. The \emph{Fermi}  source was also detected by EGRET as 3EG J1300-4406 and it is listed as an affiliation 
in the  first year AGN catalogue with no further information.
The \emph{Rosat} detection is quite strong and
is localized 4.2 degrees South West of 
the radio galaxy Centaurus A; it has a radio counterpart in the SUMSS
survey with a 36 cm flux of around 44 mJy. The X and radio detection together with the
high galactic latitude  indicate that   1FGL J1304.3-4352 has
an extragalactic origin and hence a tentative AGN classification.

{\bf 1FGL J1823.5-3454}. This source was previously detected by Enstein and ASCA but with large error boxes.
Within the \emph{Rosat} positional uncertainty lies the radio source NVSS J182338-345412, detected at  20, 
36 and possibly  6 cm \citep{wright} with a flux of  132, 
155 and 148 mJy respectively; it is also characterized by a flat radio spectrum \citep{mahony1}. Despite the location on the galactic plane 
the detection in radio, X and gamma- ray frequencies as well 
as the flat radio spectrum hints at an AGN seen through the galactic plane.

{\bf 1FGL J1227.9-4852}. This \emph{Rosat} source coincides positionally with XSS J12270-4859 classified as a cataclysmic variable and detected up to high energies 
by \emph{INTEGRAL} \citep{bird}.  Its classification has been questioned recently by \citet{martino} as the broad band characteristics 
suggest that it might be
a low mass X-ray binary system. Both  types of systems are still to be proven emitters of MeV-GeV photons and so the association 
is still uncertain. We note however that   XSS J12270-4859 is by far the brightest source in the \emph{Fermi} error box
and as such this \emph{Rosat} association still deserves some  attention.

{\bf 1FGL J1643.5-0646}. The counterpart of this \emph{Rosat} object is most likely the galaxy 2MASX J16432892-0646190, which is still unclassified;
it has a counterpart in the NVSS survey with a 20 cm flux of 28 mJy. 
No other information is available for this source  so that it is difficult to assess what type of AGN 
it may be without optical follow up observations. We note, however, that the source is listed in the sample of \emph{Fermi}  AGN affiliations
with a HSP SED class, which suggests that it is probaly a BL Lac candidate.

{\bf 1FGL J0838.6-2828}. Close to this \emph{Rosat} detection  we find the XMM Slew object XMMSL1 J083842.9-282657 
 with a 0.2-12 keV flux of 6.9$\times$10$^{-12}$ erg cm$^{-2}$ s$^{-1}$;the two objects are possibly
associated given that the distance between the two is compatible with the  error boxes of the two detections.
Within the more refined XMMSlew position we do not find any radio counterpart for this object which is located 
close to the galactic plane: both these circumstances exclude an extragalactic nature for this source and hence
suggests an uncertain classification.

{\bf 1FGL J0051.4-6242}. The \emph{Rosat} source has a hard X-ray counterpart in a bright\emph{Swift}-XRT source identified as RBS 119; 
this object is classified both in NED and  SIMBAD as a BL Lac object. The\emph{Swift}-XRT spectrum
is a  simple power law absorbed by the galactic  column density (N$_H$=1.7$\times$10$^{20}$ cm$^{-2}$) 
and  with  a photon index $\Gamma$= 2.5$\pm0.1$;  
the 0.2-12 keV flux is  5.8$\times$10$^{-12}$ erg cm$^{-2}$ s$^{-1}$. The source is reported in the SUMSS survey
with a 36 cm flux of 43 mJy. Also this source is listed in the first year  AGN catalogue as an affiliation but with no further information.
 
{\bf 1FGL J0131.2+6121}. Within the \emph{Rosat} error box lies  87GB 012752.4+610507, known in radio to exhibit 
a relativistic one-sided 
jet  and also  to show a slightly negative or close to zero spectral
index \citep{ribo}.  Indeed the source flux at 6 and 20 cm frequencies is 22 and 19 mJy respectively \citep{gregory}. 
The source has  also been reported as a radio loud  AGN  both by \citet{brinkman} and  \citet{laurent}. 
The optical spectrum displays a featureless continuum
heavily absorbed at shorter wavelengths \citep{marti}. All this observational evidence points to a blazar
interpretation for this source, most likely of the BL Lac class.

{\bf 1FGL J2056.7+4938}. The \emph{Rosat} position is compatible with a \emph{Swift}/XRT source also listed in the  XMM Slew Survey 
as XMMSL1 J205642.7+494004. This source is also 
associated to an \emph{INTEGRAL} object, IGR J20569-4940, first reported by \citet{krivonos} in their all-sky hard 
X-ray survey and then listed in the fourth IBIS catalogue\citep{bird}.  The X-ray and radio properties of this X/gamma-ray object
are fully discussed  in \citet{landi}:
in X-rays the source is relatively  bright  and variable (flux in the range 8-19$\times$10$^{-12}$ erg cm$^{-2}$ s$^{-1}$) 
while in radio it has a flat  spectrum  and is radio loud. It has been proposed by \citet{paredes} as  a microquasar candidate given its location close to the Galactic plane, but a 
blazar classification is more likely given the characteristics of the broad band emission. 

{\bf 1FGL J2146.6-1345}. This \emph{Rosat} object has  a counterpart only in the NVSS survey with a 20 cm flux of 22.5 mJy. 
No other infomation is available for this source except that it is located at high galactic latitudes; this together with the radio and
X-ray detections strongly suggest that it is most likely an AGN. Here too, we  note that the source is listed 
in the sample of \emph{Fermi} AGN affiliations  with a HSP SED class, again suggestive of  a BL Lac candidate.

{\bf 1FGL J1544.5-1127}. This source has a detection in the XMMSlew survey (XMMSL1 J154439.8-112806/XMMSL1 J154439.4-112754) and has also been obsered 
by\emph{Swift}/XRT. The XMMSlew survey 
reports two detections at different epochs with
a significantly different flux of 3.5 and 11$\times$10$^{-12}$ erg cm$^{-2}$ s$^{-1}$. The XRT spectrum  is well 
fitted with an absorbed power law having $\Gamma$=1.5 $\pm$0.1
and a galactic column density of 12.5$\times$10$^{20}$ cm$^{-2}$; in this case the 0.2-12 keV flux is around 
4$\times$10$^{-12}$ erg cm$^{-2}$ s$^{-1}$, 
i.e. close to that found in the first  XMMSlew survey detection. The source is clearly variable in X-rays but has no detection so far in radio 
despite coverage by the NVSS of this sky region.
The location of the source at high galactic latitude suggests an extragalactic origin, but the lack of a radio counterpart is intriguing 
and follow-up optical observations are necessary  to establish the nature of this X-ray source.

{\bf 1FGL J0848.6+0504}. Once again the \emph{Rosat} source has an XMMSlew counterpart (XMMSL1 J084840.1+050617) with a 0.2-12 keV flux of 
3.4$\times$10$^{-12}$ erg cm$^{-2}$ ss$^{-1}$; it is associated to the radio source FIRST J084839.6+050618, which has a quite low 20 cm flux of  around 2 mJy and is
the counterpart  of SDSS J084840.20+050611.9, classified as a 
galaxy in NED. Taken all together the source properties suggest that we are dealing here too with an AGN.

{\bf 1FGL J2329.2+3755}.  Also this \emph{Rosat} object has   a counterpart in the radio band having 
6, 20 and 92 cm fluxes of 22, 16 and 19.8 mJy respectively, 
suggestive of a flat spectrum source \citep{vollmer}.  The location at high galactic latitude suggests an extragalactic origin 
further confirmed by the radio and  X-ray detection. The source  is in fact listed 
in the sample of \emph{Fermi} AGN affiliations  with a HSP SED class, again indicative  that it may be another  BL Lac.

{\bf 1FGL J1926.8+6153}. This \emph{Rosat} object has an association with an XMM Slew source, XMMSL1 J192650.6+615446, which is identified with
 87GB 192614.4+614823 a well known radio source which is still unclassified. 
The XMMSlew  catalogue reports a 0.2-12 keV flux of 2.3$\times$10$^{-12}$ erg cm$^{-2}$ s$^{-1}$.
The source has also  a detection by\emph{Swift}/XRT. The XRT spectrum is well fitted by an absorbed
power law having $\Gamma=2.6$ $\pm0.4$ and  0.2-12 keV flux of 5.8$\times$10$^{-12}$ erg cm$^{-2}$ s$^{-1}$ the measured  absorption N$_H$=5.2$\times$10$^{20}$ at cm$^{-2}$ is purely galactic.
The X-ray  flux is clearly variable while 
the various radio detections provide evidence for  a flat spectrum source as found by Specfind 
(see also \citealt{jackson}).  The location of the source at  high galactic latitudes together with 
the X-ray and radio properties suggest  an  AGN classification. Indeed the source, 
listed among the \emph{Fermi} AGN affiliations,
is optically classified as a BL Lac and further characterized as a HSP object. 
  
{\bf 1FGL J2042.2+2427}.  Once again this  \emph{Rosat} source has an XMMSlew counterpart (XMMSL1 J204206.1+242653)  associated to the galaxy 2MASX J20420606+2426518, classified in NED as a BL Lac. The  0.2-12 keV flux is 1.7$\times$10$^{-12}$ erg cm$^{-2}$ s$^{-1}$;
it is radio loud with 6 and 20 cm  fluxes of 52 and 70 mJy respectively
\citep{griffith2}, and hence can be defined as a flat  spectrum  source
(see also \citealt{reich} and \citealt{jackson}. It is listed as an affiliation in the \emph{Fermi} AGN catalogue where it
is optically classified as a BL Lac and further characterized as a HSP object.

{\bf 1FGL J0841.4-3558}.  The \emph{Rosat} source is probably the X-ray counterpart of the star  HIP 42640 (Simbad name) of spectral type F2V,
unlikely to be a gamma-ray emitter in the \emph{Fermi} catalogue. This could well be a chance positional alignment, one of which is expected given the number of objects in the sample analysed in this work.

{\bf 1FGL J1910.9+0906}.   This \emph{Rosat} source is most likely  associated to the supernova remnant  G043.3-00.2 (also known as Was49B)
The X-ray source is slightly extended  and encompasses almost the entire extension of the remnant. Supernovae are also found 
to be associated in some numbers with GeV emission \citep{abdo1} although it is not clear at the moment how many SNR
not hosting a pulsar are \emph{Fermi} emitters.

{\bf 1FGL J0054.9-2455}.  This X-ray source  has a radio association in the NVSS with an object having a 20 cm flux of 24 mJy.
The source has been associated to  the UV excess source  2MASS J00544675-2455291,  which displays a continuous spectrum with  no lines.
In Simbad it is catalogued  as a white dwarf, but the radio emission and the high latitude location suggests that it 
could also be an extragalactic object. Indeed,  the source is present in the \emph{Fermi} AGN catalogue as an affiliated object having  no optical class
but a  HSP SED. This source also appears in the recent catalogue of \emph{Fermi} detections above 100 GeV \citep{neronov}. The above information suggests that it might be  another  BL Lac.

{\bf 1FGL J1933.3+0723}. This source, located close to the galactic plane, has a radio detection at 6 and 20 cm with a flux of 94 and 104 mJy
\citep{becker}, which implies a flat spectrum source.
No other information is available, but the radio properties and the X-ray emission  could be taken as evidence that it is another case of an AGN
behind our galaxy. 

{\bf 1FGL J1553.5-3116}. This \emph{Rosat} object has a detection in both the NVSS and in the SUMSS with a 20 and 36 cm flux of 156 and 139 mJy which
indicates a flat radio spectrum. The high latitude location as well as the X-ray and radio emission clearly indicate that this is an AGN.
Indeed the source, listed among the \emph{Fermi} AGN affiliations 
is optically classified as a BL Lac  and further characterized as a HSP object.   

{\bf 1FGL J1841.9+3220}. In this case the \emph{Rosat} source is also detected  by XRT on\emph{Swift}, which allows the 
location uncertainty to be restricted and the X-ray spectral characteristics to be studied. The smaller XRT error box allows an identification 
with the radio source RGB J1841+323. 
The XRT spectrum  has a good fit with an absorbed power law having $\Gamma$ in the range 2.1-2.5
and a galactic column density of 8.4$\times$10$^{20}$ cm$^{-2}$; the 0.2-12 keV flux is in the range 1.6-1.9$\times$10$^{-12}$ erg cm$^{-2}$ s$^{-1}$.
The available 6 and 20 cm  fluxes of 14 and 20 mJy \citep{laurent} suggests that the radio spectrum might be flat;
RGB J1841+323 has also been reported as a radio loud AGN first  
by \citet{brinkman} and later by \citet{laurent}. Finally it is listed in the Seoul National University Bright Quasar
Survey \citep{lee}, which is the base for the NED classification as a QSO candidate and hence a probable blazar type object. The source 
is listed among the \emph{Fermi} AGN affiliations with the same association proposed here and with a HSP SED class,  which indicates that it may be
a BL Lac rather than a Flat Spectrum Radio Quasar.

{\bf 1FGL J1419.7+7731}. This \emph{Rosat} detection is also a radio emitter with a 20 cm flux of 8 mJy. An optical observation of the source
indicates that it is a  weak point-like object with  an  extremely blue continuum \citep{zickgraf}; the source is also listed in the  
Million QSO catalogue (http://quasars.org/milliquas.htm). Again the  high latitude location, radio and X-ray emission  
and optical properties indicate that it is an AGN.

{\bf 1FGL J2323.0-4919}. This \emph{Rosat} source is still unidentifed and has no detection in radio to date. At a distance of around 0.7 arcmin, 
i.e. outside the \emph{Rosat} error circle we find 
an  XMMSlew survey source, XMMSL1 J232254.4-491624 which has a positional uncertainty of around 5 arcsec and 
a 0.2-12 keV flux of 2.6$\times$10$^{-12}$ erg cm$^{-2}$ s$^{-1}$. Within the XMM Slew  error circle 
there is also a radio source listed in the SUMSS with a 36 cm flux of  28.3 mJy. Although the error circles of the two X-ray sources do not match,
it is still possible that the \emph{Rosat} and XMM Slew detections are the same.
This source is also listed between the \emph{Fermi} AGN affiliations but is wrongly identified with the  galaxy APMUKS(BJ) B232010.42-493502.4, which is
not associated to  the \emph{Rosat} and/or XMMSlew detections. The source is located at high galactic latitudes, is likely an X-ray emitter 
and possibly also a radio source which all together suggest an AGN nature.
 
{\bf 1FGL J0223.0-1118} Here too the \emph{Rosat} object has an association to a XMM Slew source, XMMSL1 J022314.7-111735,  which is identified with
NVSS J022314-111737, a  radio object not yet optically classified; it is associated to  the galaxy 6dF J022314.3-111738 at redshift 0.042, but the 
optical spectrum is of too poor quality to allow a proper classification. 
The XMMSlew  catalogue reports a 0.2-12 keV flux of 1.97$\times$10$^{-12}$ erg cm$^{-2}$ s$^{-1}$, while the NVSS provides a radio detection of 14 mJy
at 20 cm. The source is clearly  an AGN in the local universe.  

\section{Discussion}
The first result of this work is that a number
of likely X-ray counterparts to \emph{Fermi} sources have been found. Statistically there should only be around one chance alignment which is probably that of 1FGL J0841.4-3558 given the type of the X-ray source. 
The majority of the associations are of extragalactic nature while only 2 or 3 cases (a SNR, a binary system and
maybe a microquasar) are galactic. Of the extragalactic objects many are BL Lac or BL Lac candidates, 
i.e. objects that  are expected to have GeV emission. In other cases the source maybe radio loud or radio flat, 
characteristics that are often common to AGN emitting in the \emph{Fermi} band (see Table 2 and the previous section). 

Some information on the nature of the \emph{Fermi} sources can also  be gained from the GeV properties 
as reported in the \emph{Fermi} catalogue;
for each source of interest here Table 2 provides the \emph{Fermi} 1-100 GeV flux, power law photon index,  
curvature index and  variability index.For example both curvature and variability index can be used to discriminate between source types: 
as shown in figure 11 of \citet{abdo1}
one can clearly separate the pulsar branch located at large curvature and small variability indices from the blazar 
branch  which is found  at large variability  and small curvature indices.
Using these parameters and following the broad division adopted by Abdo and co-workers, we 
conclude that  all extragalactic objects in Table 1 are compatible with being blazars;  the few exception 
are 1FGL J1942.7+1033,  1FGL J1227.9-4852 and 1FGL J1910.9+0906 which are border line objects. The last two are indeed
non blazar type objects while 1FGL J1942.7+1033 despite its location in the diagram is most likely  a BL LAC 
given its broad band properties (see previous section).

In the following discussion we concentrate only on those objects that are likely to be extragalactic and so 
exclude the 3 sources which are galactic or spurios (1FGL J1227.9-4852,  1FGL J1910.9+0906 and 1FGL J0841.4-3558)
but leave 1FGL J2056.7+4938 as it could well be a blazar behind the galactic plane.
To go deeper in our understanding of the nature of the \emph{Rosat}-\emph{Fermi} associations, we can use the gamma-ray photon index 
to discriminate between BL Lac and FSRQs.  From figure 12 in \citet{abdo2}, we see the latter have a lower limit to the \emph{Fermi} photon index of 2, while that for BL Lac objects peak around this value with a range from about 1.2 to 2.7. 
Of the 27 extragalactic sources 16 have a photon index below this critical value and so must be strongly suspected to be BL Lacs. Furthermore there are another 7 objects which have a steeper spectrum but have already been optically classified as BL Lacs therefore in total we have at least 23 objects which probably belong to this class.
Similarly, the intensity of the GeV emission suggests a preference for 
BL Lac objects among our sample, since the log of the \emph{Fermi} flux reported in table 2 is always  below -8.0.
(see figure 10 in \citealt{abdo2}). Finally if we plot the 0.1-2.4  X-ray flux  versus the  flux density  at 20 cm for those objects which have both values,   
we find that their location 
in this diagram is again in the region populated by BL Lac objects (see figure 5 of \citealt{abdo2}).

Thus it seems that the cross correlation using the \emph{Rosat} bright source catalogue tends to select associations with \emph{Ferm}i sources that are BL Lac type AGN.  To test this finding we can use the same statistical correlation method but on the sample of \emph{Fermi} objects
that are classified in the first catalogue as blazars (bzq and bzb). There are 573 such sources in the \emph{Fermi} list, 
with almost the same number of objects in each of the two clases (51\% BL Lac).
Again the correlation is strong with 181 associations within 300 arcseconds. Of these objects, the overwhelming majority (137) are with those already identified as BL Lac sources. The selection effect towards BL Lac when using the  \emph{Rosat} bright source catalogue  is likely related to the spectral energy distribution (SED)
of these objects compared to FSRQ. In the widely adopted scenario of blazars, a single population of 
high-energy electrons in a relativistic jet radiate from the radio/FIR to 
the UV- soft X-ray by the synchrotron process and at higher frequencies by inverse Compton scattering of soft-target 
photons present either in the jet (synchrotron self-Compton [SSC] model), in the surrounding
material (external Compton [EC] model), or in both (\citet{ghisellini}
and references therein). Therefore a strong signature of the Blazar nature 
of a source is a double peaked structure in the SED, with the synchrotron component 
peaking anywhere from Infrared to X-rays and the
inverse Compton extending up to GeV or even TeV gamma-rays. Among blazars, 
BL Lacertae objects are the sources with the highest variety of synchrotron peak frequencies, 
ranging from the IR-optical to the UV-soft-X bands (called Low or High energy peak BL Lacs, respectively, see \citet{padovani}). 
The X-ray selection discussed herein should favour objects peaking at high energies, i.e. in the X-ray band: indeed  9 objects discussed in the previous 
section and affiliated to the \emph{Fermi} AGN catalogue, are also classified as high synchrotron peaked AGN.
Given their high synchrotron peak energies, which flag the presence of high energy electrons, 
these extreme BL Lacs are also good candidates for TeV emission as the Compton peak is expected in this energy range. One source in our sample, 
1FGL J0648.8+1516, has already been 
detected by VERITAS in the TeV range while another, 1FGL 0054.9-2455, has been seen above 100 GeV. The interest in these extreme TeV blazars is driven  by the possibility of obtaining information both on the acceleration processes of charged particles in relativistic flows and on the intensity  of the extragalactic background light which absorbs the flux from high energy sourxces \citep{mank}.

\section{Conclusions}
We have shown that, as expected, there is a strong correlation between the \emph{Fermi} survey source 
list and the \emph{Rosat} All Sky Survey  Bright Source Catalogue, finding that there should be about 60 sources common to both lists.
By placing a maximum correlation distance of 160 arcseconds in order to minimise chance associations we have a sample of 30 objects in which only $1 \pm 1$ should be by chance alignment.
We can use the \emph{Rosat} error box to help find optical counterparts 
for these sources, which in all  cases is sufficiently small to allow the identification of one single object.  Most of the associations are of extragalactic nature, with only a few being galactic.
We have also shown that this cross correlation analysis appears to preferentially select BL Lac objects. These X-ray selected objects  are often HSP BL Lacs and are thereffore good candidates for TeV emission.
Clearly only optical spectroscopy of the \emph{Rosat} counterparts can confirm this suggestion, but cross correlation with other catalogues may provide a different selection criterion and hence different source types.

\section*{Acknowledgments}
This research has been partially supported by ASI contract I/008/07/0
This analysis has made use of the HEASARC archive which is a service of the Laboratory for High Energy Astrophysics (LHEA) at NASA/ GSFC
and the High Energy Astrophysics Division of the Smithsonian Astrophysical Observatory (SAO). It has also used
the NASA/IPAC Extragalactic Database (NED) which is operated by the Jet Propulsion Laboratory, California Institute of Technology, under contract with the National Aeronautics and Space Administration and the  SIMBAD database, operated at CDS, Strasbourg, France

\end{document}